\documentclass[english,notitlepage,longbibliography,superscriptaddress,lnofootinbib]{revtex4-1}
\usepackage[T1]{fontenc}
\usepackage[latin9]{inputenc}
\usepackage{color,times,ulem}
\usepackage{mathrsfs}
\usepackage{amsmath}
\usepackage{amssymb}
\usepackage{graphicx}
\usepackage{hyperref}
\hypersetup{colorlinks,linkcolor={blue},citecolor={blue},urlcolor={red}}  

\makeatletter

\providecommand{\tabularnewline}{\\}

\usepackage{babel}

\makeatother

\usepackage{babel}
\begin{document}
\title{Enhancement of non-Gaussianity and nonclassicality of photon added
displaced Fock state: A quantitative approach}
\author{Priya Malpani$^{1}$, Kishore Thapliyal$^{2}$, J. Banerji$^{3}$,
Anirban Pathak$^{1,*}$ \\
{\small{}$^{1}$Jaypee Institute of Information Technology, A-10,
Sector-62, Noida, UP-201307, India}\\
{\small{}$^{2}$Joint Laboratory of Optics of Palack\'{y} University
and Institute of Physics of CAS, Faculty of Science, }\\
{\small{}Palack\'{y} University, 17. listopadu 12, 771 46 Olomouc,
Czech Republic}\\
{\small{}$^{3}$Physical Research Laboratory, Navrangpura, Ahmedabad
380009, India}}
\begin{abstract}
Non-Gaussian and nonclassical states and processes are already found
to be important resources for performing various tasks related to
quantum gravity and quantum information processing. The effect of
non-Gaussianity inducing operators on the nonclassicality of quantum
states has also been studied rigorously. Considering these facts,
a quantitative analysis of the nonclassical and non-Gaussian features
is performed here for photon added displaced Fock state, as a test case, using a
set of measures like entanglement potential, Wigner Yanese skew information,
Wigner logarithmic negativity and relative entropy of non-Gaussianity.
It is observed that photon addition (Fock parameter) significantly
increases the amount of nonclassicalty and non-Gaussianity for small
(large) values of the displacement parameter, which decreases both
the quantum features monotonically. In this respect, the role of Fock
parameter is found to be more prominent and stronger compared to photon
addition. Finally, the dynamics of Wigner function under the effect
of photon loss channel is used to show that only highly efficient
detectors are able to detect Wigner negativity.
\end{abstract}
\maketitle

\section{Introduction \label{sec:Introduction}}

Quantification of non-Gaussianity and nonclassicality have drawn considerable
attention in the recent past. This attention is well deserved as with
the advent of quantum information science, it has been realized that
nonclassicality and quantum non-Gaussianity are two of the most important
facets of quantum world, which lead to quantum advantage \citep{dowling2003quantum}.
Quantification problems have been approached in various ways. However,
no unique measure of nonclassicality or non-Gaussianity can be built.
We begin with the definition of nonclassicality and non-Gaussianity
before discussing their significance in quantum technology and the requirement
to quantify these resources.

The notion of nonclassicality arises form the non-positive value of the
Glauber-Sudarshan $P$ function \citep{sudarshan1963equivalence,glauber1963coherent}
\begin{equation}
\rho=\int d\left(P\left(\alpha\right)\right)\left|\alpha\right\rangle \left\langle \alpha\right|.\label{eq:P-fun}
\end{equation}
In other words, a state that cannot be represented in terms of a statistical
mixture of coherent states is known as a nonclassical state. In 
recent years, the importance of nonclassical states has been enhanced
considerably as in the domain of meteorology and information processing
\citep{pathak2018classical}, a major demand for the enhancement of
the performance of the devices appeared and it is established that
the desired enhancement is possible by using quantum resources. Nonclassical
states and properties are found to be essential for obtaining quantum
advantage and thus, they appeared as the basic building blocks for
quantum or quantum-enhanced devices having advantages over the corresponding
classical versions. Here, it may be noted that the nonclassical states
not only lead to the detection of gravitational waves \citep{aasi2013enhanced},
they are also found essential for quantum teleportation \citep{bennett1993teleporting,bouwmeester1997experimental,brassard1998teleportation},
quantum key distribution \citep{bennett1984quantum,acin2006bell,srikara2020continuous},
quantum computation \citep{nielsen2010quantum,pathak2013elements},
etc. As nonclassicality leads to quantum advantage, attempts have
been made to quantify the amount of nonclassicality and the quantum
advantage it can provide. Specifically, in 1987, a distance based
measure was introduced by Hillery \citep{hillery1987nonclassical},
but there were many computational difficulties associated with that.
Following that, Mari et al. \citep{mari2011directly} gave a measure
based on trace norm distance. In 1991, Lee introduced a measure for
the quantification of nonclassicality known as nonclassical depth
\citep{lee1991measure} (for a brief review see \citep{miranowicz2015statistical}).
The studies based on these measures revealed that non-Gaussianity
inducing operators, e.g., photon addition and other quantum state
engineering operations \citep{agarwal2013quantum,malpani2019lower}
can induce and/or enhance the amount of nonclassicality in an arbitrary
quantum state. 

In addition, we may note that non-Gaussian states can be defined as
the states which cannot be defined in terms of a probabilistic mixture
of Gaussian states \citep{genoni2013detecting,kuhn2018quantum}. On
the other hand, all the Gaussian states can be defined by the first
two moments, which means that the mean and covariance matrix of the
Gaussian states provide us their complete information. We can define
a complex hull $\mathcal{G}$ with the classical probability distribution
$P_{{\rm cl}}\left(\lambda\right)$ 
\begin{equation}
\rho=\int d\left(P_{{\rm cl}}\left(\lambda\right)\right)\left|\psi_{G}\left(\lambda\right)\right\rangle \left\langle \psi_{G}\left(\lambda\right)\right|\label{eq:Gauss}
\end{equation}
in the Hilbert space $\mathcal{H}$. This set contains all the Gaussian
states and some non-Gaussian states. Interestingly, the non-Gaussian
states obtained as the statistical mixtures of Gaussian states in the
form (\ref{eq:Gauss}) have limited applications due to its origin
in classical noise \citep{franzen2006experimental,kuhn2018quantum}.
On the other hand, quantum non-Gaussian states $\rho$ in $\mathcal{H}$
are the states which do not belong to the complex hull $\rho\notin\mathcal{G}$.
It is worth to study the quantification of quantum non-Gaussianity
in such a state as these states can be used as more robust
resources compared to the Gaussian states (\citep{menicucci2006universal,adesso2009optimal,albarelli2018resource}
and references therein). Specifically, there are no-go theorems limiting
the applications of Gaussian operations and Gaussian states in entanglement
distillation \citep{fiurasek2002gaussian}, quantum error correction
\citep{niset2009no}, quantum computing \citep{menicucci2006universal} and
quantum bit commitment \citep{magnin2010strong}. Use of non-Gaussian
operations is known to provide advantages in computation \citep{menicucci2006universal},
communication \citep{chatterjee2020quantifying}, metrology \citep{adesso2009optimal},
etc. These set of promising applications have motivated witnesses
of non-Gaussianity \citep{genoni2013detecting}, resource theory of
non-Gaussianity \citep{albarelli2018resource}, and a set of measures
of quantum non-Gaussianity (\citep{lachman2019faithful,lachman2021quantum,hlouvsek2021direct}
and references therein). Further, it may be noted that traditionally
quantum gravity and quantum information processing have evolved as
two independent branches, but in the recent past, various significant
and exciting results connecting these two fields have been reported
(see \citep{howl2021non} and references therein). Specifically, in
\citep{howl2021non}, it is shown that non-Gaussianity present
in matter can be used to detect the signature of quantum gravity as
only quantum theory of gravity can lead to non-Gaussianity. Although,
in what follows, we will discuss non-Gaussianity of field, this exciting
connection between non-Gaussianity of matter and quantum gravity has
also motivated us to study how the amount of non-Gaussianity varies
with different state parameters of photon added displaced Fock state
(PADFS). 

There are a couple of reasons for choosing PADFS as the test bed for
the present study on the measures of nonclassicality and non-Gaussianity.
Firstly, at different limits, this state reduces to a set of well-known
quantum states having a large number of applications. Further, in
previous works of some of the present authors, nonclassical properties
\citep{malpani2019lower} and phase properties \citep{malpani2019quantum}
of PADFS were studied. There it was observed that non-Gaussianity
inducing operators significantly affect the signatures of nonclassicality
viewed through different witnesses of nonclassicality. 

Motivated by the above facts, we set ourselves the task to quantify
the nonclassicality and quantum non-Gaussianity of PADFS using some
measures, namely linear entropy potential, skew information based
measure, relative entropy based measure of non-Gaussianity and Wigner
logarithmic negativity. Here, it is also worth stressing that the
skew information based measure and Wigner logarithmic negativity fail
to detect nonclassicality in some quantum states and are thus categorized
as quantifiers of nonclassicality. However, our study on the skew
information based measure is restricted to the pure states, where we
can use it as a reliable measure of nonclassicality. Highly efficient
detectors are required to detect Wigner negativity. This motivated
us to quantify the nonclassicality and non-Gaussianity based on Wigner
negativity over the photon loss channel, which also models inefficient
real detectors used in experiments. 

The rest of the paper is organized as follows. In Section \ref{sec:Photon-added-displaced},
the properties of PADFS and its limiting cases are discussed. In the
next section, we report nonclassicality measures based on entanglement
potential, skew information based measure, Wigner logarithmic negativity,
and non-Gaussianity based measure. A comparative study of measures
of nonclassicality and non-Gaussianity is performed in Section \ref{sec:Comparative-study-of}.
In Section \ref{sec:Decoherence:-Noisy-Wigner}, the dynamics of the
Wigner function and Wigner logarithmic negativity for PADFS over photon
loss channel are reported. Finally, the paper is concluded in Section
\ref{sec:Result-and-discussion}. 

\begin{figure}
\begin{centering}
\includegraphics{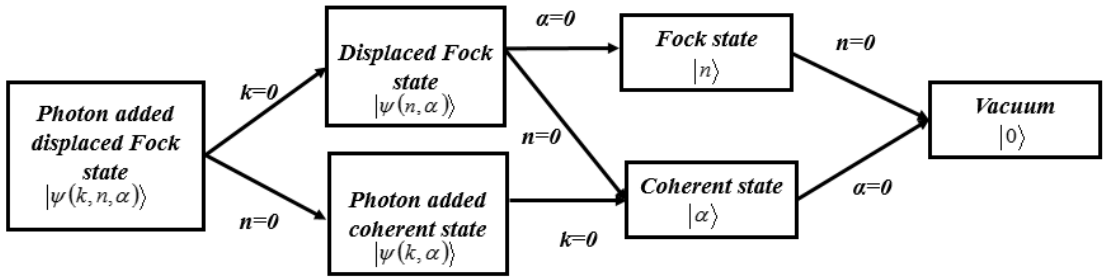}
\par\end{centering}
\caption{\label{fig:limiting cases-1} Various limiting cases of PADFS.}
\end{figure}

\section{Photon added displaced Fock state and its Wigner function \label{sec:Photon-added-displaced}}

Displaced Fock state (DFS) is analytically expressed as $D\left(\alpha\right)\left|n\right\rangle $,
where $D(\alpha)$ is displacement operator and $|n\rangle$ is the
Fock state. It has been studied very rigorously in the past
\citep{de1990properties,de2005alternative,de2006nonlinear,dodonov2005decoherence,lvovsky2002synthesis,marchiolli2004engineering}
because of the fact that this state is nonclassical but its special
case (displaced vacuum state which is known as coherent state and
is obtained for $n=0$) is classical. In the present work, the state
of interest is PADFS \citep{malpani2019lower} which is obtained by
the application of the creation operator (non-Gaussianity inducing
operator) on the DFS \citep{de1990properties}. In the previous works,
some of the present authors have reported the nonclassical and phase
properties of PADFS through various witnesses \citep{malpani2019lower,malpani2019quantum}.
However, neither non-Gaussianity was discussed nor the nonclassicality
present in PADFS was quantified. This will be attempted here,
but before we do that, it will be apt to introduce the state of our
interest and to specifically state the motivation for this particular
choice of state.

The analytic expression for $k$ photon added DFS can be given as
\begin{equation}
\left|\psi\left(k,n,\alpha\right)\right\rangle =\sum_{m=0}^{\infty}C_{m}\left(n,k,\alpha\right)\left|m+k\right\rangle ,\label{eq:PADFS-1}
\end{equation}
where 
\begin{equation}
\begin{array}{lcl}
C_{m}\left(n,k,\alpha\right) & = & N\exp\left[-\frac{\left|\alpha\right|^{2}}{2}\right]\frac{\sqrt{n!\left(m+k\right)!}}{m!}\alpha^{m-n}L_{n}^{m-n}\left(\left|\alpha\right|^{2}\right)\end{array}\label{eq:cm}
\end{equation}
with normalization factor
\begin{equation}
\begin{array}{lcl}
N & = & \left(\sum_{m=0}^{\infty}\exp\left[-\left|\alpha\right|^{2}\right]\left|\frac{\sqrt{n!\left(m+k\right)!}}{m!}\alpha^{m-n}L_{n}^{m-n}\left(\left|\alpha\right|^{2}\right)\right|^{2}\right)^{-\frac{1}{2}}.\end{array}\label{eq:norm}
\end{equation}
Here, $n$ is the Fock parameter, $L_{n}^{l}$ is 
an associated Laguerre polynomial, and $\alpha$ is the displacement
parameter. In principle, PADFS, described as above, can be generated experimentally,
and the schematic diagram of the setup to be used to generate PADFS
was reported by us in our previous work \citep{malpani2019lower}.
Apart from the experimental realizability, the PADFS is of interest
because in different limits, it reduces to different important and
useful quantum states (including vacuum, Fock, Coherent, photon added
coherent states, etc.). In Fig. \ref{fig:limiting cases-1}, the limiting
cases of PADFS are explicitly shown. In what follows, we obtain the
signature of nonclassicality and non-Gaussianity present in PADFS
using Wigner function as that can reveal signatures of both the features
(i.e., non-Gaussianity and nonlcassicality).

\begin{figure}
\begin{centering}
\includegraphics[scale=0.6]{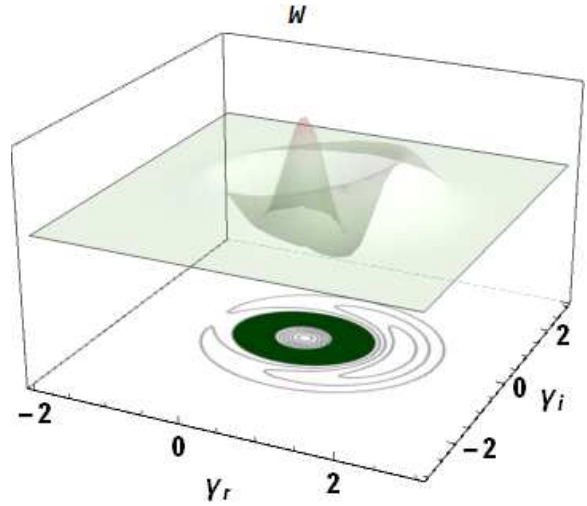}
\par\end{centering}
\caption{\label{fig:Wigner-1} (Color Online) Wigner function for the single
PADFS with Fock parameter $n=1$ and displacement parameter $\alpha=0.5$.
In the contour plot at the bottom, the dark part represents the negative region whereas the contour lines indicate
the positive region of the Wigner function.}
\end{figure}

One can define Wigner function \citep{wigner1932quantum} of a quantum
state with density matrix $\rho$ in the coherent state representation as
\begin{equation}
W\left(\gamma,\gamma^{\star}\right)=\frac{2}{\pi^{2}}\exp\left[-2\left|\gamma\right|^{2}\right]\int d^{2}\lambda\langle-\lambda|\rho|\lambda\rangle\exp\left[-2\left(\gamma^{\star}\lambda-\gamma\lambda^{\star}\right)\right].\label{eq:wigner-def}
\end{equation}
We have calculated (see Appendix \ref{sec:Appendix-A}) the
analytical expression for Wigner function of PADFS as
\begin{equation}
W\left(\gamma,\gamma^{\star}\right)=\frac{2\left|N\right|^{2}}{n!\pi}\exp\left[-2\left|\gamma-\alpha\right|^{2}\right]\left[\sum_{t=0}^{n}S_{tt}P_{tt}+2{\rm Re}\sum_{t=1}^{n}\sum_{t^{\prime}=0}^{t-1}S_{t^{\prime}t}P_{t^{\prime}t}\right],\label{eq:Wigner}
\end{equation}
where 
\[
P_{t^{\prime}t}=\left(-1\right)^{k+t}\left(k+t^{\prime}\right)!\eta^{t-t'}L_{k+t'}^{t-t'}\left(\left|\eta\right|^{2}\right)
\]
with $\eta=-\alpha+2\gamma$ and 
\[
S_{t^{\prime}t}=\left(\begin{array}{c}
n\\
t^{\prime}
\end{array}\right)\left(\begin{array}{c}
n\\
t
\end{array}\right)\alpha^{\star\left(n-t^{\prime}\right)}\alpha^{\left(n-t\right)}
\]
are polynomials depending upon photon addition $\left( k\right)$ and Fock parameter $\left( n\right)$,
respectively.

The existence of nonclassical and non-Gaussian features in the state
chosen here can be studied in various ways. Notice that for $k=n=0$
the Wigner function (\ref{eq:Wigner}) has Gaussian form (as it corresponds
to the Wigner function of coherent state). The polynomial in the product
of this Gaussian function (i.e., $S_{t^{\prime}t}P_{t^{\prime}t}$)
has a degree more than one for $k\neq0$ and/or $n\neq0$. For example,
in the Wigner function of photon added coherent state for $k\neq0=n$,
we have $S_{t^{\prime}t}=1$ and $P_{t^{\prime}t}=\left(-1\right)^{k}k!L_{k}\left(\left|\eta\right|^{2}\right)$
responsible for the non-Gaussian nature of the corresponding state.
Similarly, in the case of DFS, $P_{t^{\prime}t}=1$ and $S_{t^{\prime}t}$
is polynomial of degree more than one for $n>0$. Further, the non-zero
value of displacement parameter can influence the non-Gaussian behavior
iff either $k\neq0$ or $n\neq0$. Thus, one can easily conclude qualitatively
that the polynomial in the product with a Gaussian factor due to photon
addition and non-zero Fock parameter leads to non-Gaussianity.

One may further study the nonclassical and non-Gaussian behavior of
PADFS for different choice of state parameters by plotting the Wigner
function. Specifically, the negative values of Wigner function indicate
the presence of nonclassicality. On top of that, according to Hudson's
theorem \citep{hudson1974wigner,soto1983wigner}, a pure quantum state
having positive Wigner function is always Gaussian. This infers that
the negative values of Wigner function (\ref{eq:Wigner}) witness
the non-Gaussianity in the state. Keeping this point into consideration,
we have obtained the Wigner function shown in Fig. \ref{fig:Wigner-1},
which clearly reveals the nonclassical and non-Gaussian behavior.
Especially, the existence of both these features is clearly observed
through the negative values of the Wigner function. Note that the
Wigner function only witnesses the presence of nonclassicality and
non-Gaussianity, but does not help us to quantify the amount of nonclassicality
and non-Gaussianity. In what follows, we quantify both these features
using different measures of nonclassicality and non-Gaussianity.

\section{Measures of nonclassicality and non-Gaussianity\label{sec:Measures}}

Photon addition is a frequently used operation in quantum state engineering
and it is often used to induce non-Gaussianity. Consequently, it is
usually considered as a non-Gaussianity inducing operator. Further,
it can also be used to introduce nonclassicality by hole burning \citep{escher2004controlled,gerry2002hole}.
Here, we aim to study the variation of the amount of nonclassicality
and non-Gaussianity present in PADFS with the state parameters (i.e.,
the number of photons added $(k)$, the Fock $(n)$ and displacement
$(\alpha)$ parameters). The relevance of the study underlies in the
fact that such a study may help us to identify the suitable state
parameters for an engineered quantum state which is to be used to
perform a quantum computation, communication or metrology task that
requires nonclassical and/or non-Gaussian states. For the quantification
of nonclassicality, the closed form analytical expressions
of an entanglement potential (to be referred to as linear entropy potential),
skew information based measure, and Wigner logarithmic negativity
are obtained in this section. Further, Wigner logarithmic negativity
and relative entropy of non-Gaussianity are computed as measures of
non-Gaussianity. In what follows, we would briefly introduce these
measures and explicitly show how the amount of nonclassicality and
non-Gaussianity quantified by these measures vary with the state parameters.

\subsection{Entanglement potential: Linear entropy potential }

To begin the discussion on the measures of nonclassicality, we must
note that there are several nonclassicality measures (e.g., nonclassical
depth \citep{lee1991measure}, distance based measures \citep{hillery1987nonclassical}),
but each has its own limitation(s) (for a quick review see \citep{miranowicz2015statistical}).
In 2005, Asboth \citep{asboth2005computable} introduced
a new measure of nonclassicality, which quantifies the single-mode
nonclassicality using the fact that if a single-mode nonclassical
(classical) state is inserted from an input port of a beamsplitter
(BS) and vacuum state is inserted from the other port then the output
two-mode state must be entangled (separable). Consequently, a measure
of entanglement can be used to indirectly measure the nonclassicality
of the input single-mode state (the state other than the vacuum state
inserted in the BS). However, there exist many quantitative
measures of entanglement and any of those can be used to measure the
nonclassicality of the input single-mode state. When a measure of
entanglement is used to measure the single mode nonclassicality using
this approach, corresponding entanglement measure is referred to as
the entanglement potential in analogy with the terminology used by
Asboth. Specifically, if we use concurrence (linear entropy) to measure
the entanglement of the post-BS two-mode output state in an effort
to quantify the nonclassicality of the single-mode input state, then the nonclassicality measure is called concurrence potential
(linear entropy potential) \citep{miranowicz2015statistical}. In
what follows, we quantify the nonclassicality of PADFS using Asboth's
approach considering linear entropy potential as the entanglement
potential.

\begin{figure}
\centering{}%
\begin{tabular}{cc}
\includegraphics[scale=0.6]{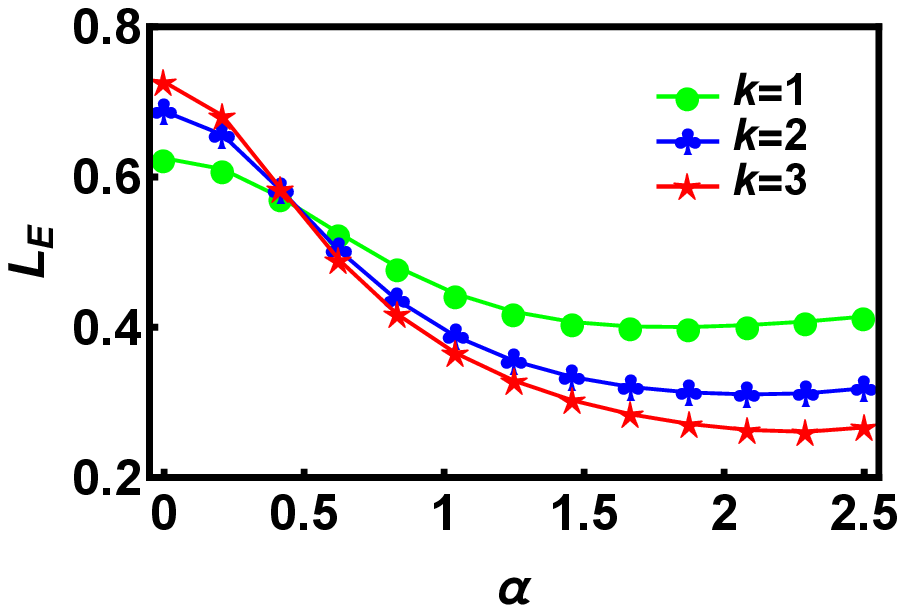}  & \includegraphics[scale=0.6]{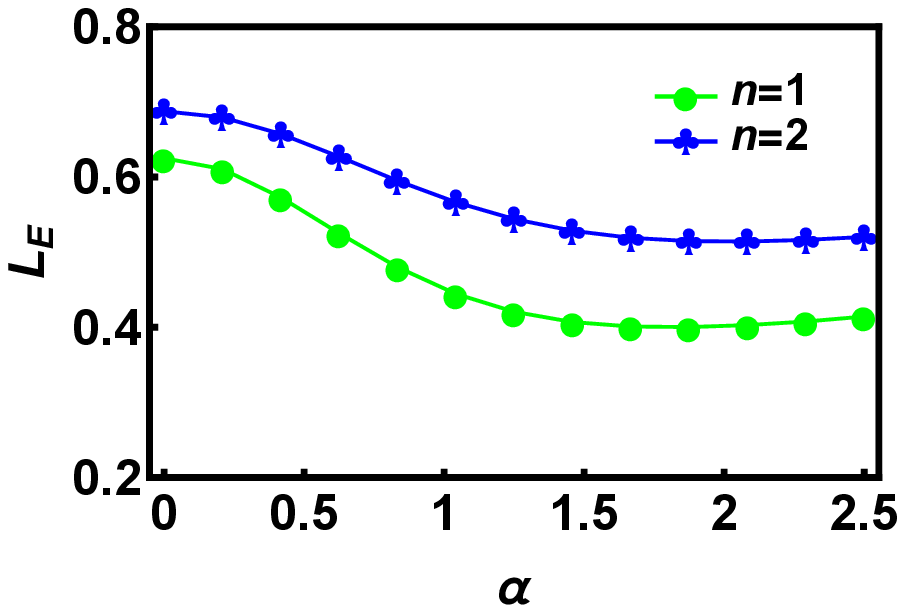}\tabularnewline
(a)  & (b)\tabularnewline
\end{tabular}\caption{\label{fig:Linear Entropy}(Color Online) Linear entropy for PADFS
as a function of $\alpha$, (a) for different number $\left( k\right)$ of photon addition
and $n=1$ and (b) for different Fock parameter $\left( n\right)$ and $k=1$.}
\end{figure}

We know that the linear entropy for a bipartite state $\rho_{AB}$
is defined in terms of the purity of the reduced subsystem as
\begin{equation}
L_{E}=1-{\rm Tr}\left(\rho_{B}^{2}\right),\label{eq:le}
\end{equation}
where $\rho_{B}$ can be obtained by taking partial trace over the subsystem
$A$. Thus, $L_{E}=0$ $(1)$ for a separable (maximally entangled)
state and a non-zero value for an entangled state in general \citep{Meher2018,wei2003maximal}.
To compute it, we would require the post-BS state $\rho_{AB}$ that
will originate if PADFS and vacuum states are mixed at a BS. To begin
with, we may note that the post-BS transformation can be
understood in terms of Hamiltonian $H=-\hbar\theta(a^{\dagger}b+ab^{\dagger})$,
where $a^{\dagger}$, $b^{\dagger}$ are the creation operators and
$a$ and $b$ are annihilation operators for the two input modes. Also, we have chosen $\theta=\pi/4$ here for a symmetric-BS parameter. 
The expression for the two mode state having Fock state at one port and
vacuum at another port can be written as
\begin{equation}
|n\rangle\otimes|0\rangle=|n,0\rangle\stackrel{{\rm BS}}{\longmapsto}\frac{1}{2^{n/2}}\sum_{j=0}^{n}\sqrt{^{n}C_{j}}\,\,|j,\,n-j\rangle.\label{eq:inout_fock}
\end{equation}
This equation can be used to obtain post-BS state $\rho_{AB}=|\phi\rangle_{AB}\,_{AB}\langle\phi|$
that originates on the insertion of PADFS and vacuum state from the
input ports of a BS with 
\begin{equation}
\begin{array}{lcl}
{  {{  |\phi\rangle_{AB}}}} & {  {{  =}}} & {  {{  |\psi(k,n,\alpha)\rangle\otimes|0\rangle}{  \stackrel{{\rm BS}}{\longmapsto}}{  \sum\limits _{m=0}^{\infty}C_{m}\left(n,k,\alpha\right)}{  \sum\limits _{k_{1}=0}^{m+k}\sqrt{{m+k \choose k_{1}}}\left(\frac{1}{\sqrt{2}}\right)^{k+m}}}} {  {  {{  i^{m+k-k_{1}}}{  |k_{1},m+k-k_{1}\rangle.}}}}
\end{array}\label{two mode state}
\end{equation}
In order to calculate the expression for linear entropy, we obtain
partial trace over the subsystem $A$ in the post-BS state $\rho_{AB}$
and use Eq. (\ref{eq:le}) to yield an analytic expression for linear
entropy as
\begin{equation}
\begin{array}{lcl}
{{  L_{E}}} & {  =} & {  {  1}{  -}{{  \sum\limits _{m,m^{\prime},r=0,}^{\infty}C_{m}\left(n,k,\alpha\right)C_{m^{\prime}}^{\star}\left(n,k,\alpha\right)C_{r}\left(n,k,\alpha\right)C_{m+r-m^{\prime}}^{\star}\left(n,k,\alpha\right)}}}\\
 & \times & {{  \sum\limits _{k_{1}=0}^{m+k}\sqrt{{m+k \choose k_{1}}}}}{{  \sqrt{{m^{\prime}+k \choose k_{1}}{r+k \choose r-m^{\prime}+k_{1}}{m-m^{\prime}+r+k \choose r-m^{\prime}+k_{1}}}}{  \left(\frac{1}{2}\right)}^{{  m+r+2k}}}.
\end{array}\label{eq:LE-PADFS}
\end{equation}

We illustrate the variation of the amount of nonclassicality with
respect to the state parameters in Fig. \ref{fig:Linear Entropy}
which quantifies the nonclassicality of PADFS through linear entropy
potential. Nonclassicality decreases with increase in the displacement
parameter as for $\alpha=0$ the state reduces to Fock state, the most
nonclassical state. The value of the displacement parameter is significant
in controlling the outcome of photon addition. Specifically, Fig.
\ref{fig:Linear Entropy} (a) shows that up to a certain value of
displacement parameter $\alpha$, nonclassicality increases with photon
addition while an opposite behavior is observed after a certain value
of $\alpha$. In what follows, we will refer to this particular
value of $\alpha$ as $\alpha_{{\rm inversion}}$ as, at this value
of $\alpha$, the amount of nonclassicality based ordering of different
PADFSs get inverted. Specifically, if we make an ascending order of
single photon added, two photon added and three photon added DFS (with
the same initial Fock parameter), based on the amount of nonclassicality
quantified by linear entropy potential, then from Fig. \ref{fig:Linear Entropy}
(a) we can see that for $\alpha<\alpha_{{\rm inversion}}$ the ordered
sequence is single photon added DFS, two photon added DFS, three photon
added DFS, but the sequence is opposite for $\alpha>\alpha_{{\rm inversion}}.$
Interestingly, $\alpha_{{\rm inversion}}$ depends on all the state
parameters, such as ${  L}_{{  E}}^{k=1}={  L}_{{  E}}^{k=2}$
for $\alpha_{{\rm inversion}}=0.45$ while ${  L}_{{  E}}^{k=2}={  L}_{{  E}}^{k=3}$
for $\alpha_{{\rm inversion}}=0.42$ when Fock parameter is $n=1$.
Further, from Fig. \ref{fig:Linear Entropy} (b), it is easy to observe
that with an increase in the Fock parameter the nonclassicality is
found to be increased, and the performance is better than in the case of photon addition. In other words, photon addition is
an effective tool for enhancing nonclassicality for the small values of $\alpha$
while higher values of the Fock parameter are more beneficial at larger
values of the displacement parameter. This can be attributed to the origin
of nonclassicality from corresponding operations as photon addition
burns a hole in the photon number distribution, which influences more
on the state with the small values of $\alpha$ due to larger probability amplitude
for vacuum in the Fock basis.

\subsection{Skew information based measure}

This skew information based measure of nonclassicality was introduced
by Shunlong Luo et al. \citep{luo2019quantifying} in 2019 and is
based on Wigner Yanase skew information \citep{wigyur1963}. In the case
of a pure state $\rho$, this is defined as
\begin{equation}
N\left(\rho\right)=\frac{1}{2}+\left\langle a^{\dagger}a\right\rangle -\left\langle a^{\dagger}\right\rangle \left\langle a\right\rangle .\label{eq:N}
\end{equation}
$N\left(\rho\right)$ has a clear meaning as the quantum coherence $\rho$ with respect
to annihilation operator $a$, and it is relatively easy to compute.
In the case of PADFS, we can obtain
\begin{equation}
{{  N\left(\rho\right)}{  =}{  \frac{{  1}}{{  2}}}{  +}}{  {{  \sum\limits _{m=0}^{\infty}}{  \left|C_{m}\left(n,k,\alpha\right)\right|}^{{  2}}}}{  \left(m+k\right)}-\left|{  {{  \sum\limits _{m=0}^{\infty}}{  C_{m}\left(n,k,\alpha\right)}}{{  C_{m+1}^{\star}\left(n,k,\alpha\right)}{  \sqrt{m+k+1}}}}\right|^{2}.\label{eq:N-PADFS}
\end{equation}

This measure is based on averages, which takes numerical value 
$\frac{1}{2}$ for coherent state (classical) and 
$n+\frac{1}{2}$ for $n$ photon Fock state (most nonclassical).
Thus, any state $\rho$ with $N(\rho)>\frac{1}{2}$ is nonclassical.
However, this is mentioned in \citep{luo2019quantifying} as only
the sufficient criterion of nonclassicality in general as it fails
for some Gaussian mixed states. Further, $N\left(\rho\right)$ also
quantifies the quantum metrological power associated with the given
quantum state. 

\begin{figure}
\centering{}%
\begin{tabular}{cc}
\includegraphics[scale=0.6]{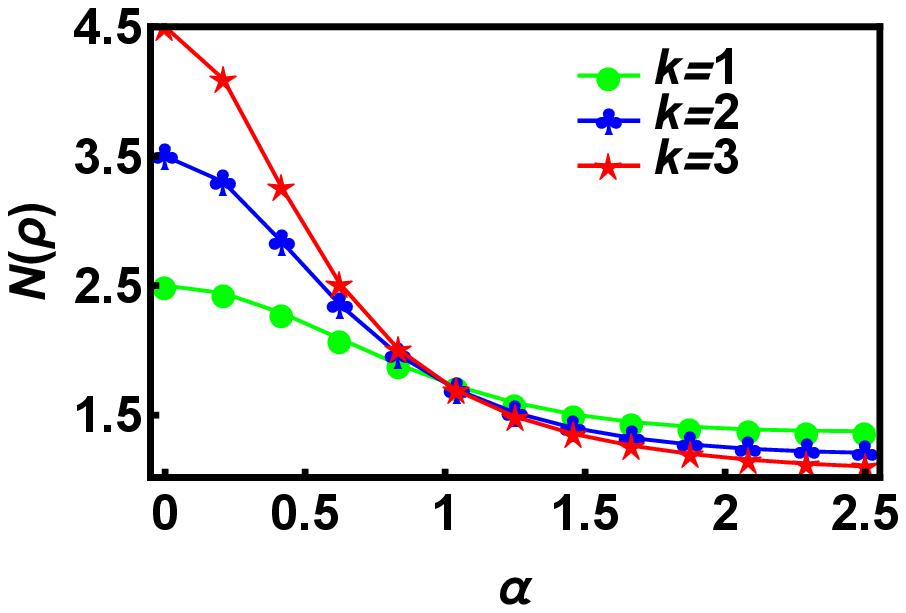}  & \includegraphics[scale=0.6]{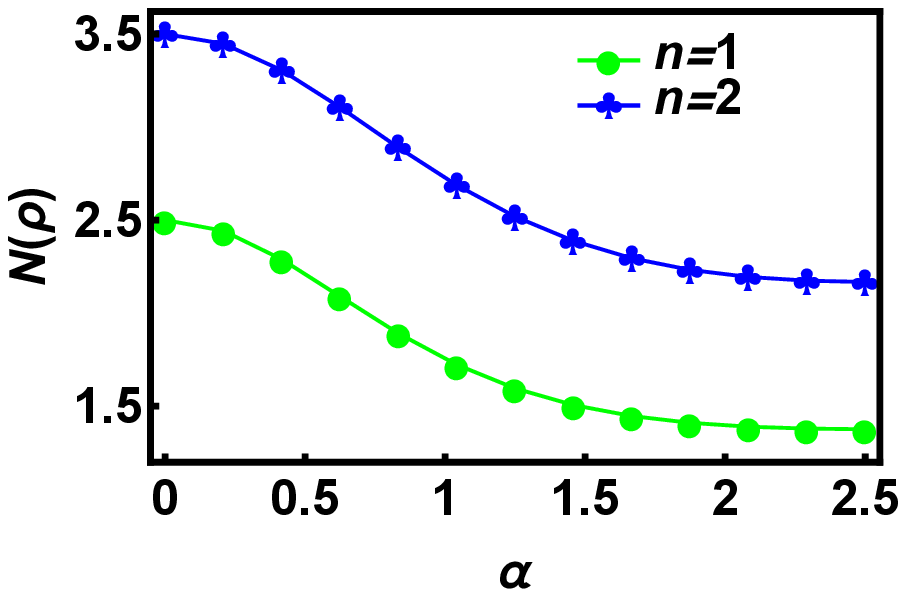}\tabularnewline
(a)  & (b)\tabularnewline
\end{tabular}\caption{\label{fig:PASDFS}(Color online): Nonclassicality quantifier for
$k$ PADFS having Fock parameter $n$ with respect to $\alpha$ (a)
for different photon addition $\left( k\right)$ Fock parameter $n=1$, (b) for different
Fock parameter $\left( n\right)$ and photon addition $k=1$.}
\end{figure}

From the obtained expression of this skew information based measure
(\ref{eq:N-PADFS}) we observed that nonclassicality cannot be enhanced
by increasing the displacement parameter. Further, Fig. \ref{fig:PASDFS}
(a) shows that with an increase in photon addition, nonclassicality
increases up to a certain value of displacement parameter and decreases
thereafter. This is analogous to our earlier observations in the context of
linear entropy potential. Interestingly, the value of $\alpha_{{\rm inversion}}$
is different, i.e., it is larger than that in the corresponding case for
linear entropy potential. This is consistent with some of the present
authors' earlier observations that nonclassical states cannot be strictly
ordered on the basis of the amount of nonclassicality as the nonclassicality
measures are not monotone of each other \citep{miranowicz2015statistical}.
This point may be further clarified with a specific example. If we
consider linear entropy potential as a measure of nonclassicality,
then single photon added DFS is more nonclassical than two photon
added DFS with $\alpha=0.75$, but two photon added DFS would appear
to be more nonclassical with $N(\rho)$ as the measure of nonclassicality,
indicating that the validity of any such ``amount of nonclassicality''-based
ordering is restricted to the specific measure used. Moreover, Fig.
\ref{fig:PASDFS} (b) shows that nonclassicality increases with increase
in the Fock parameter, as noted earlier (Fig. \ref{fig:Linear Entropy}
(b)) in the case of linear entropy.

\subsection{Wigner logarithmic negativity \label{subsec:Wigner-log-neg}}

We have already mentioned in Section \ref{sec:Photon-added-displaced}
that the negative values of Wigner function are signatures of nonclassicality.
This motivated to use the volume of the negative part of the Wigner
function as a quantifier of nonclassicality \citep{kenfack2004negativity}.
Recently, further extending this idea, a measure of nonclassicality,
named Wigner logarithmic negativity, is introduced as \citep{albarelli2018resource}
\begin{equation}
\mathscr{W}=\log_{2}\left(\int d^{2}\gamma\left|W\left(\gamma\right)\right|\right),\label{eq:WLN-1}
\end{equation}
where the integration is performed over whole phase space. Interestingly,
in the resource theory of quantum information, it has been noted that
$\mathscr{W}$ also quantifies the amount of non-Gaussianity present
in the state as the negative values of Wigner function (cf. Fig. \ref{fig:Wigner-1})
also witness the non-Gaussianity of PADFS. 

The integration in Eq. (\ref{eq:WLN-1}) is performed numerically
using Wigner function (\ref{eq:Wigner}) to study the effect of different
parameters on the Wigner logarithmic negativity of PADFS. The Wigner logarithmic negativity shows similar behavior with all
the parameters as that for linear entanglement potential and skew
information based measure. Specifically, any increase in the displacement
(Fock) parameter only reduces (increases) nonclassicality and non-Gaussianity,
whereas with increase in photon addition the Wigner logarithmic negativity
increases (decreases) for small (large) displacement parameter (cf.
Fig. \ref{fig:Linear Entropy}). Among the photon addition and Fock
parameter, the latter performs better than the former in enhancing
nonclassicality and non-Gaussianity. Although, a Gaussian operation,
namely displacement operation, is not expected to increase non-Gaussianity
of DFS but it plays very significant role in determining the non-Gaussianity
of PADFS.

\begin{figure}
\begin{centering}
\begin{tabular}{cc}
\includegraphics[scale=0.5]{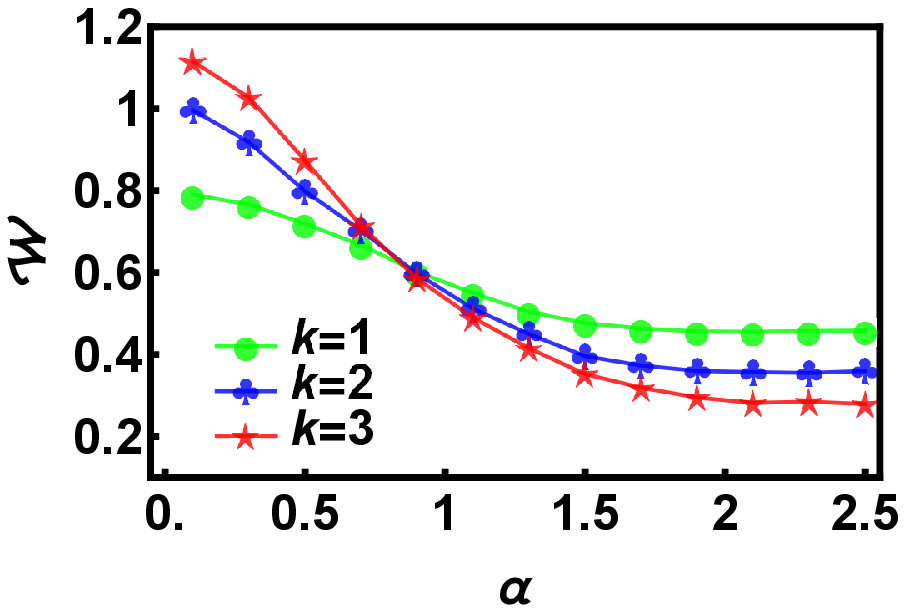} & \includegraphics[scale=0.5]{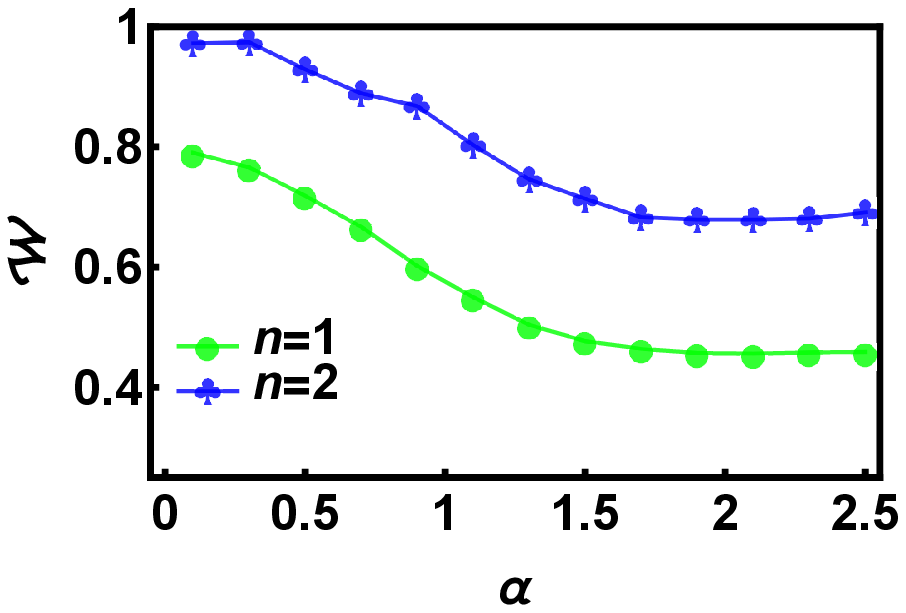}\tabularnewline
(a)  & (b)\tabularnewline
\end{tabular}
\par\end{centering}
\caption{\label{fig:nonclassical-vol} (Color Online) Variation of Wigner logarithmic
negativity with displacement parameter (a) for different values of
photon addition $\left( k\right)$ and $n=1$ and (b) for different values of Fock parameter $\left( n\right)$ and
single photon addition $k=1$. }
\end{figure}

\subsection{Measure of non-Gaussianity}

A number of criteria to quantify non-Gaussianity have been reported
in literature \citep{albarelli2018resource,ivan2012measure,kuhn2018quantum,park2017quantifying,takagi2018convex}.
In the previous section, we have quantified non-Gaussianity using
Wigner logarithmic negativity and observed that in the case of PADFS,
variation of the amount of non-Gaussianity with the state parameters
of interest is analogous to that of nonclassicality as quantified
through different measures. This is justified as the set of Wigner
negative states is a subset of non-Gaussian states \citep{albarelli2018resource}.
In what follows, we aim to re-investigate this feature by quantifying
non-Gaussianity using relative entropy of non-Gaussianity.

\subsubsection{Relative entropy based measure of non-Gaussianity\label{subsec:n-g}}

The relative entropy of a non-Gaussian state with the set of all Gaussian
states is defined as 
\[
\delta\left[\rho\right]=S\left(\rho\shortparallel\tau_{G}\right),
\]
where the relative entropy $S\left(\rho||\Lambda\right)={\rm Tr}\left[\rho\left(\log\rho-\log\Lambda\right)\right]$,
and the reference mixed Gaussian state $\tau_{G}$ is chosen to have
the same first and second moments as that of $\rho$. However, in the
case of pure states, $S(\rho=\left|\psi\right\rangle \langle\psi|)=0$
and we can define $\delta\left[\left|\psi\right\rangle \right]=S\left(\tau_{G}\right)$,
von Neumann entropy of the reference state. 

Further, we know that a Gaussian state is fully characterized by its first
two moments and thus its covariance matrix. Specifically, the covariance
matrix $\sigma$ can be written as \citep{albarelli2018resource} 
\begin{equation}
\sigma=\left[\begin{array}{lc}
\sigma_{qq} & \sigma_{qp}\\
\sigma_{qp} & \sigma_{pp}
\end{array}\right],\label{eq:co-var}
\end{equation}
where $\sigma_{uv}=\left(\left\langle uv+vu\right\rangle -2\left\langle u\right\rangle \left\langle v\right\rangle \right)$
for position $q=\frac{a+a^{\dagger}}{\sqrt{2}}$ and momentum $p=\frac{a-a^{\dagger}}{i\sqrt{2}}$.
All the elements of covariance matrix $\sigma$ for PADFS are obtained
as 
\[
\begin{array}{lcl}
{  \left(\begin{array}{c}
{  {{  \sigma}_{{  qq}}}}\\
{{  \sigma}_{{  pp}}}
\end{array}\right)} & = & {  {{  \sum_{m=0}^{\infty}}{  \left\{ {  {\rm Re}\left(C_{m+2}\left(n,k,\alpha\right)C_{m}^{\star}\left(n,k,\alpha\right)\right)}{  \sqrt{\left(m+k+1\right)\left(m+k+2\right)}}\right.}}}\\
 & {  \pm} & {{  \left.2\left|C_{m}\left(n,k,\alpha\right)\right|^{2}\left(m+k\right)\left(m+k-1\right)-1\right\} }}\\
 & {{  \pm}} & {{  \sum_{m=0}^{\infty}\left(C_{m+1}\left(n,k,\alpha\right)C_{m}^{\star}\left(n,k,\alpha\right)\pm{\rm c.c.}\right)^{2}\left(m+k+1\right)}}
\end{array}
\]
and 
\[
\begin{array}{lcl}
{  {{  \sigma_{qp}}}} & {  {{  =}}} & {{  \sum_{m=0}^{\infty}{\rm Im}\left(C_{m+2}\left(n,k,\alpha\right)C_{m}^{\star}\left(n,k,\alpha\right)\right)\sqrt{\left(m+k+1\right)\left(m+k+2\right)}}}\\
 & {  -} & {{  \sum_{m=0}^{\infty}{\rm Im}\left(C_{m+1}^{2}\left(n,k,\alpha\right)C_{m}^{\star2}\left(n,k,\alpha\right)\right)\left(m+k+1\right)}},
\end{array}
\]
where ${\rm Re}(z)$ and ${\rm Im}(z)$ correspond to the real and
imaginary parts of the complex number $z.$ Thus, relative entropy
of non-Gaussianity in terms of the covariance matrix reduces to
\begin{equation}
\delta\left[\left|\psi\right\rangle \right]=S\left(\tau_{G}\right)=h\left(\det\left(\sqrt{\sigma}\right)\right),\label{eq:n-g}
\end{equation}
where $h\left(z\right)=\left(\frac{z+1}{2}\right)\log_{2}\left(\frac{z+1}{2}\right)-\left(\frac{z-1}{2}\right)\log_{2}\left(\frac{z-1}{2}\right).$

\begin{figure}
\begin{centering}
\begin{tabular}{cc}
\includegraphics[scale=0.6]{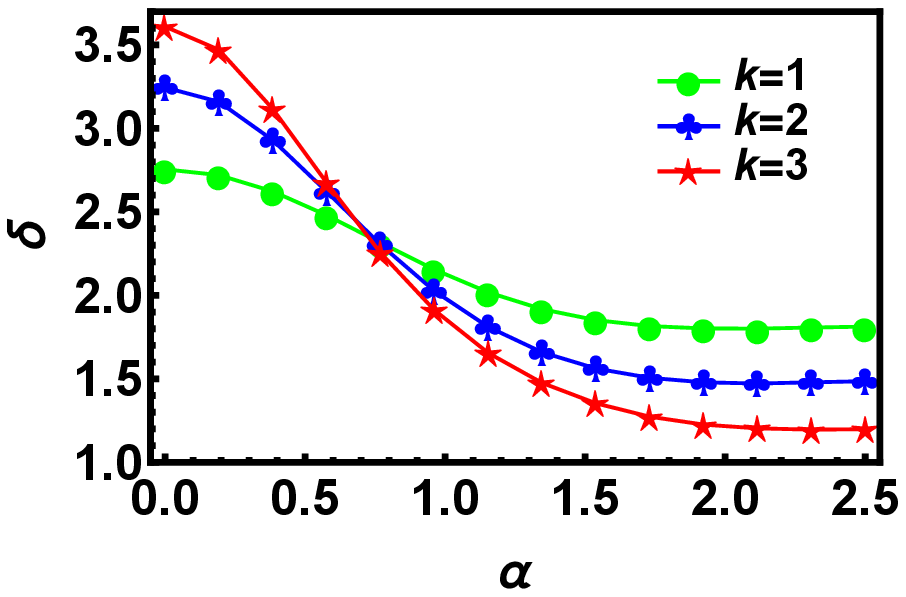}  & \includegraphics[scale=0.6]{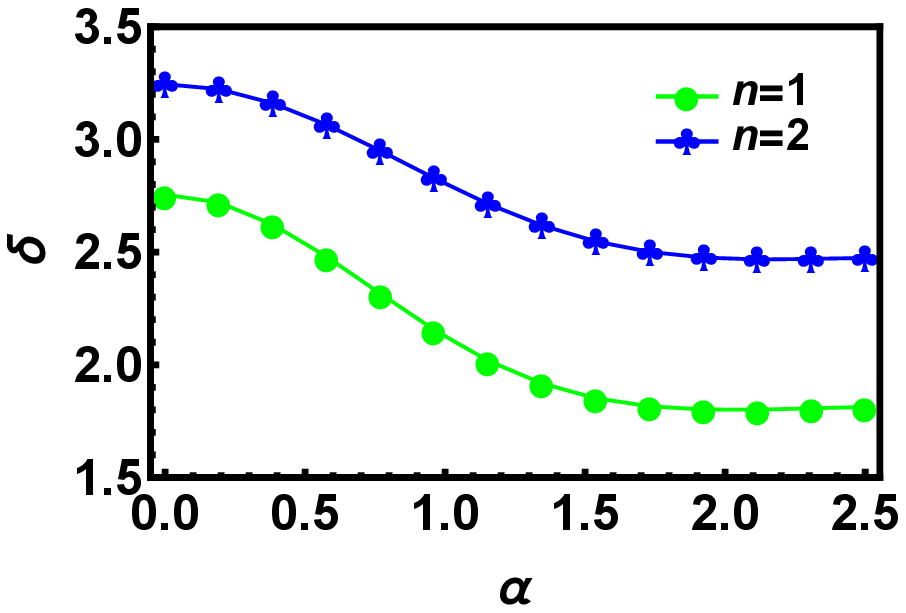}\tabularnewline
(a)  & (b)\tabularnewline
\end{tabular}
\par\end{centering}
\caption{\label{fig:Single_photon} (Color Online) Non-Gaussianity as a function
of $\alpha$ (a) for variation in photon addition with $n=1$, (b)
with respect to different Fock parameter for $k=1$. }
\end{figure}

It may be noted that the variation of the amount of non-Gaussianity
present in a single photon added coherent state (PACS) has already
been studied using different measures of non-Gaussianity (see Refs.
\citep{fu2020quantifying,genoni2013detecting,barbieri2010non}). Interestingly,
for $k=1,n=0$, PADFS reduces to PACS, and we observe that in this
specific case, variation of non-Gaussianity reported here in a more
general scenario provides similar pattern as reported in the earlier
works (cf. Fig. 2 of \citep{fu2020quantifying}) where it was observed
that non-Gaussianity reduces with $\alpha$. For a coherent state,
$\alpha^{2}$ (if $\alpha$ is considered to be real) is the average
photon number. Similarly, for PACS average photon number is $\frac{\alpha^{4}+3\alpha^{2}+1}{\alpha^{2}+1}$
and a much more complex expression of average photon number as a function
of $\alpha$ for PADFS is reported in \citep{malpani2019lower}. In
all these cases, average photon number is found to increase with $\alpha$.
For a relatively large value of $\alpha$, average photon number is
high. Consequently, addition of one or two photons or equivalently
application of non-Gaussianity inducing creation operator for once
or twice does not have much impact on non-Gaussianity or nonclassicality.
This intuitive logic explains the nature of plots (initially decreasing
with $\alpha$ and gradually approaching a line parallel to the horizontal
axis) observed here and also in earlier studies. Further, for
$\alpha\rightarrow0,$ PADFS reduces to a Fock state $|m\rangle=|k+n\rangle$.
Figs. \ref{fig:nonclassical-vol} and \ref{fig:Single_photon} clearly
show that non-Gaussianity of Fock state $|m\rangle$ increases with
$m$ which is also consistent with the earlier results. For example,
in a recent work \citep{fu2020quantifying}, non-Gaussianity has been
quantified using an uncertainty relation and it is found that non-Gaussianity
of Fock state $|m\rangle$ is $m$. Thus, addition of photon to a
Fock state will always increase non-Gaussianity, but the same is not
true in general for a superposition of Fock states. 

\begin{figure}
\centering{}%
\includegraphics[scale=0.5]{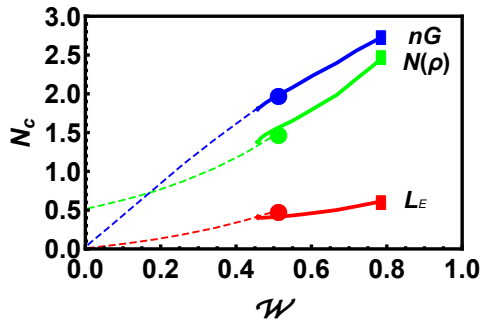}
\caption{\label{fig:parametricplot} (Color Online): Parametric plot as a function
of Wigner logarithmic negativity for various measures ($N_{c}$) for
PADFS (solid lines) with $k=1\text{, \ensuremath{n=1}}$, PACS (dashed
lines) with $k=1$, and solid circle (square) representing the values
of different measures for single photon (two photon) Fock state $\left|1\right\rangle $
$\left(\left|2\right\rangle \right)$.}
\end{figure}

\section{A comparative study of measures of nonclassicality and non-Gaussianity\label{sec:Comparative-study-of}}

So far we have studied nonclassicality and non-Gaussianity independently
with a quantitative approach, but noted that Wigner function can be
used to witness both the features. Further, photon addition in a Gaussian
state (coherent state) is known to be significant in inducing both the
features. More recently comparison of qualitative behavior of the variation of the amount of non-Gaussianity quantified using two measures, namely relative entropy based measure and Wigner logarithmic negativity, is
shown using a parametric plot \citep{albarelli2018resource}. Here, we show the amount of
nonclassicality and non-Gaussianity obtained by different measures
through a parametric plot as function of the amount of Wigner logarithmic
negativity. Specifically, the amount of nonclassicality quantified
through linear entropy potential and skew information based measure
as well as that of non-Gaussianity measured using relative entropy
based measure as function of the amount of Wigner logarithmic
negativity for PADFS (solid lines), PACS (dashed lines) and Fock state
is illustrated in Fig. \ref{fig:parametricplot}. One can verify
that the amount of nonclassicality and non-Gaussianity in PACS varies
linearly between that of the values for the single photon and vacuum.
On the other hand, due to non-zero Fock parameter in PADFS, the amount
of nonclassicality and non-Gaussianity in PADFS can be observed to
vary approximately between that of the values for the single-photon
and two-photon Fock states. The slope of the line representing the
variation of linear entropy potential is the least as the entanglement based
measure is upper bounded to unity while the rest of the measures
are not bounded.

\section{Wigner function of PADFS evolving under photon loss channel \label{sec:Decoherence:-Noisy-Wigner}}

An interaction between the quantum system and its surrounding induces
quantum to classical transition. Therefore, the observed nonclassical
and non-Gaussian features are expected to decay due to the evolution
of PADFS under lossy channel. Specifically, the temporal evolution
of a quantum state $\rho$ over lossy channel can be studied using
LGKS master equation \citep{breuer2002theory}
\begin{equation}
\frac{\partial\rho}{\partial t}=\kappa\left(2a\rho a^{\dagger}-a^{\dagger}a\rho-\rho a^{\dagger}a\right),\label{master-eq}
\end{equation}
where $\kappa$ is the rate of decay. Analogously, the time evolution
of Wigner function at time $t$ in terms of initial Wigner function
of the state evolving under lossy channel \citep{wang2011nonclassicality}
can be defined as
\begin{equation}
W(\zeta,t)=\frac{2}{T}\int\frac{\partial^{2}\gamma}{\pi}\exp\left[-\frac{2}{T}\left|\zeta-\gamma e^{-\kappa t}\right|^{2}\right]W\left(\gamma,0\right),\label{eq:noisy-wigner}
\end{equation}
where $T=1-\exp\left(-2\kappa t\right).$ The time
evolution of {Wigner function (\ref{eq:noisy-wigner})} models dissipation
due to interaction with a vacuum reservoir as well as inefficient
detectors with efficiency $\eta=1-T$.

\begin{figure}
\begin{centering}
\begin{tabular}{ccc}
\includegraphics[scale=0.4]{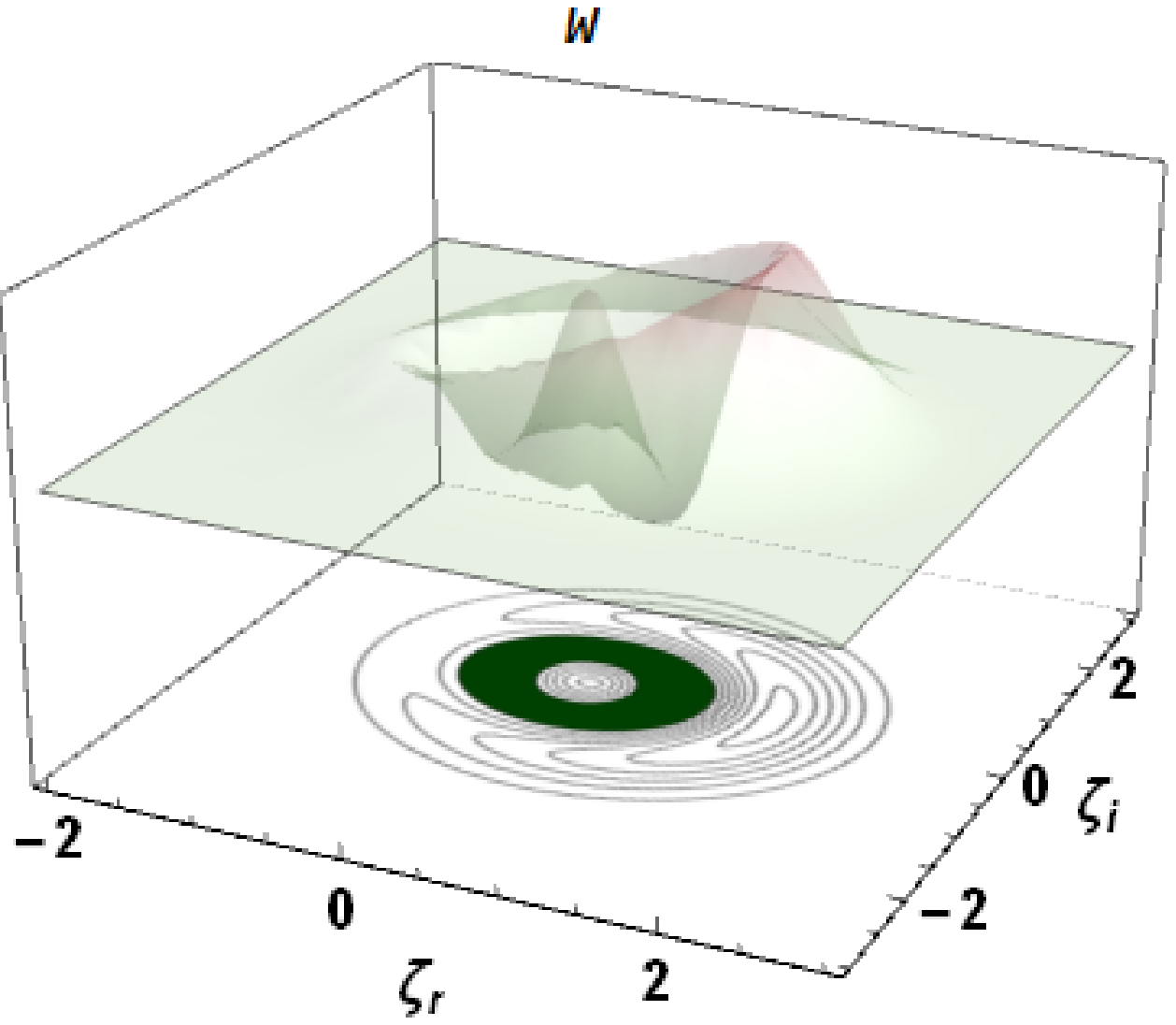}  & \includegraphics[scale=0.4]{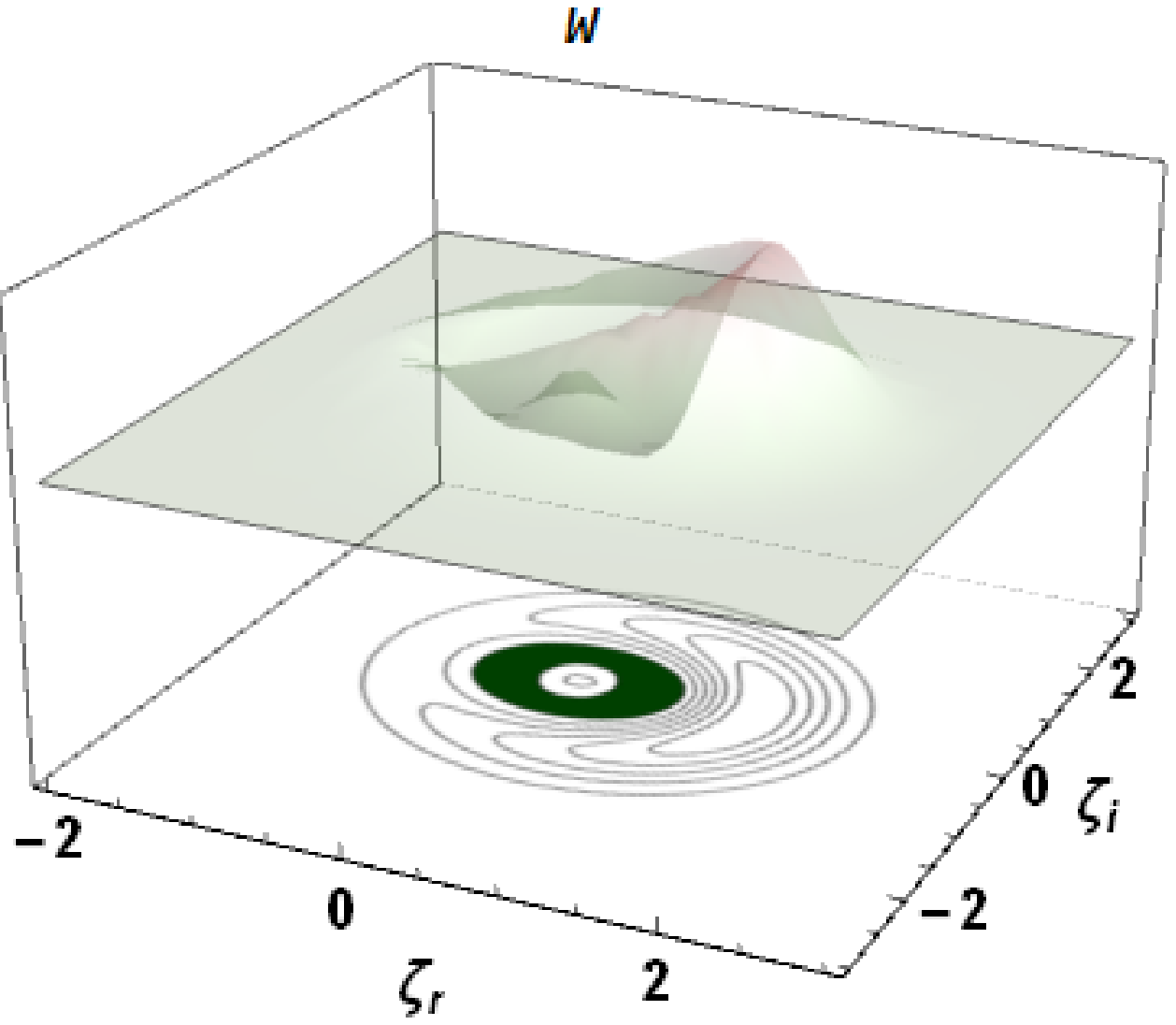}  & \includegraphics[scale=0.4]{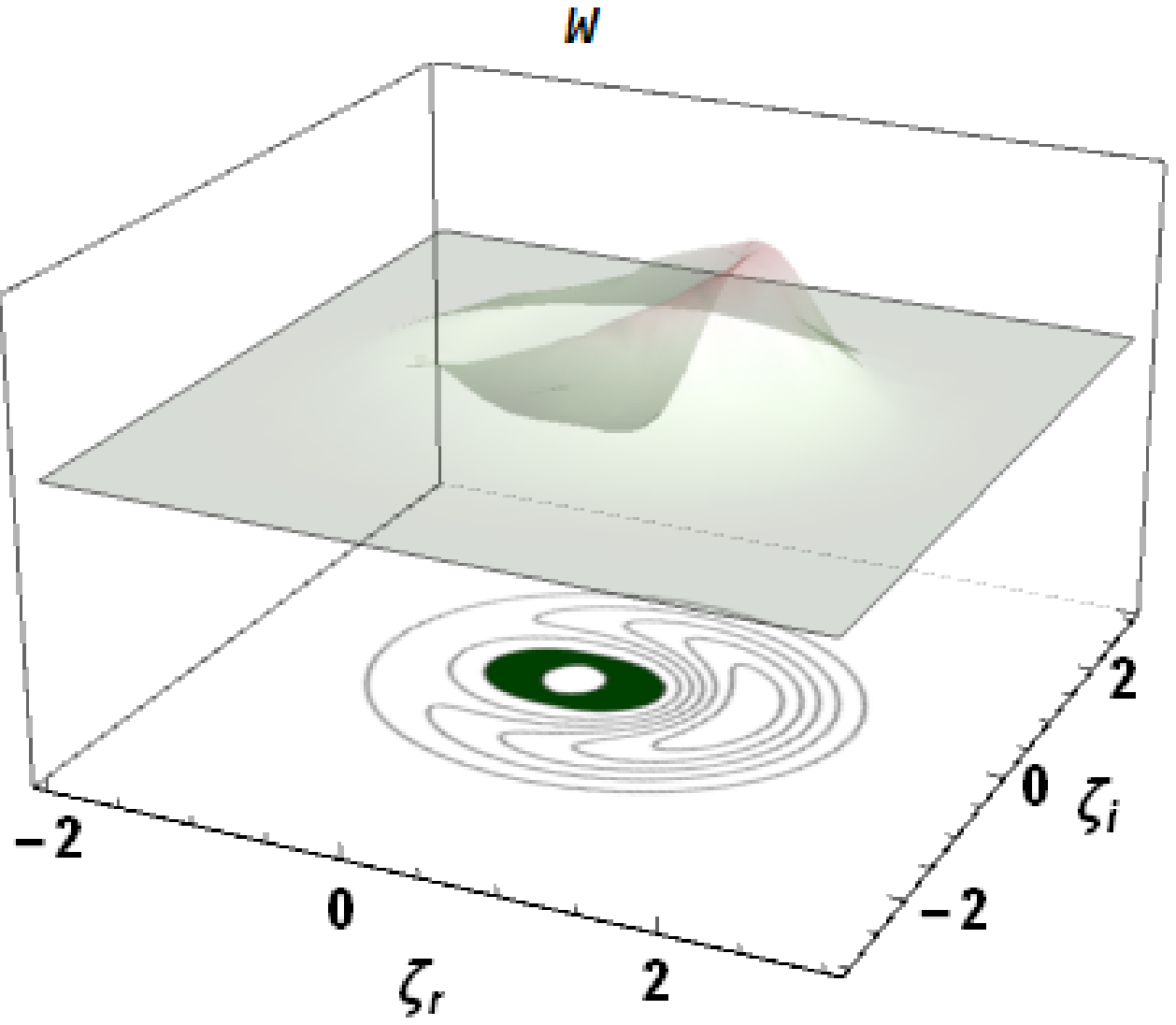}\tabularnewline
(a)  & (b)  & (c)\tabularnewline
\end{tabular}
\par\end{centering}
\caption{\label{fig:Noisy wigner}(Color Online) The Dynamics of the Wigner
function for single PADFS ($\alpha=0.5,$$n=1$) for different values
of rescaled time $\kappa t$ for (a) $\kappa t=0.1$, (b) $\kappa t=0.2$,
and (c) $\kappa t=0.25$. Here, it can be observed that with the increase
in $\kappa t$ the nonclassicality decreases. In the contour plots at the bottom, the dark part represents
the negative region and the contour lines show the positive region of the
Wigner function.}s
\end{figure}

We can clearly observe that the Wigner function at non-zero time $t$
is obtained as the convolution of the initial Wigner function. Using
the description of Wigner function in an ideal scenario, a compact
expression of Wigner function can be obtained. One can easily notice
from {Eq. (\ref{eq:noisy-wigner})} that both non-Gaussianity and nonclassicality
of PADFS cannot increase due to its temporal evolution over a photon
loss channel. This is clearly illustrated through Fig. \ref{fig:Noisy wigner}
where we can observe the negative region shrinking with increasing
values of rescaled time $\kappa t$ for a single photon added and
single Fock parameter PADFS with displacement parameter $\alpha=0.5$.
In other words, the area of dark color in the contour plots representing the
negative region is decreasing as we are increasing $\kappa t$ implying
that the quantum features are diminishing. Further, as expected from
Eq. (\ref{eq:noisy-wigner}) the quantum state loses its nonclassical
character as $t\rightarrow\infty$, and the Wigner function becomes
positive (as it reduces to that of the vacuum). In fact, the averaging
involved in the dynamics of Wigner function smooths its behavior
which in turn reduces its non-Gaussian features as well. For instance,
the peak at the origin of the phase space can be observed to be decreasing
with time in Fig. \ref{fig:Noisy wigner}. This also shows that
the present results can be used to study a non-Gaussianity witness
based on the value of the Wigner function at the origin of the phase
space \citep{genoni2013detecting}. However, it is worth mentioning
here that as the Wigner function fails to detect all the nonclassicality
(i.e., Wigner function is positive for squeezed states) and non-Gaussianity
(as non-Gaussian state with positive Wigner function is reported in
\citep{filip2011detecting}), it would be appropriate to refer to
the observed features under the effect of photon loss channel as the
lower bound on the nonclassicality and quantum non-Gaussianity in
the state. In other words, the nonclassicality and non-Gaussianity
in the PADFS measured using realistic inefficient detector may be
more than that reflected through the quantifier based on the Wigner
negativity, which we referred here as lower bound.

\begin{figure}
\begin{centering}
\begin{tabular}{cc}
\includegraphics[scale=0.5]{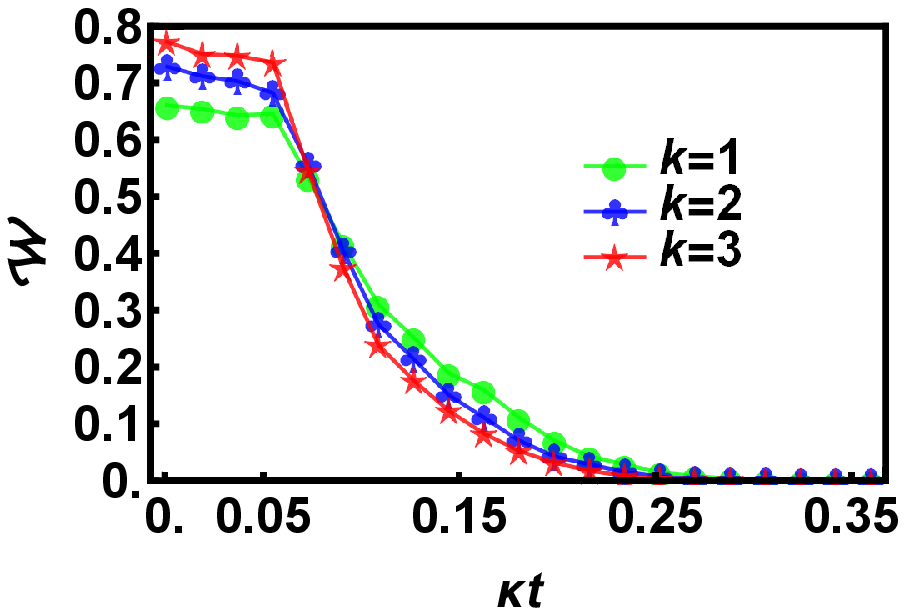} & \includegraphics[scale=0.5]{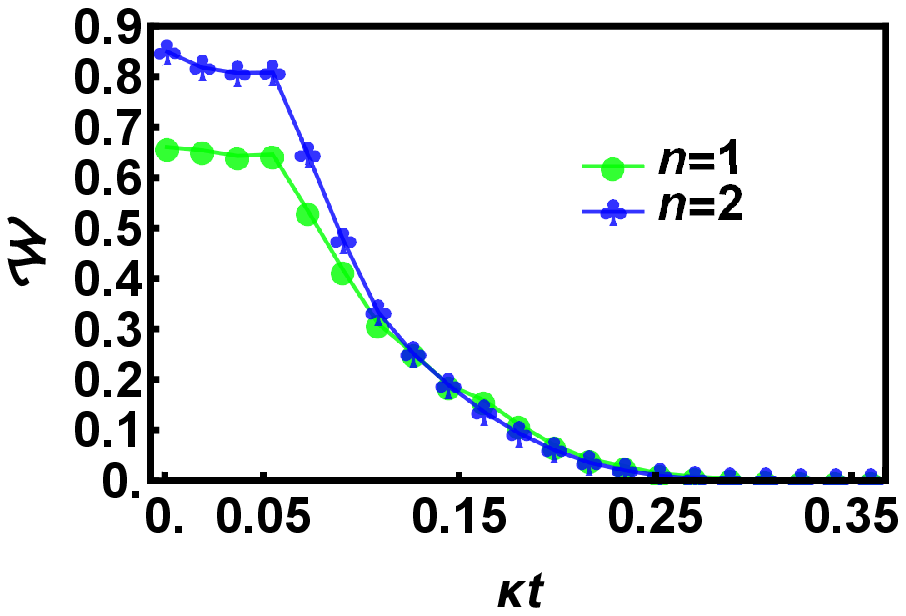}\tabularnewline
(a)  & (b)\tabularnewline
\end{tabular}
\par\end{centering}
\caption{\label{fig:wigner-log}(Color Online) The declining Wigner logarithmic
negativity of PADFS with $\alpha=0.5$ with respect to rescaled time
$\kappa t$ (a) for various photon addition and $n=1$, and (b) for
different Fock parameter and single photon addition.}
\end{figure}

To observe the above finding in a quantitative manner, the dynamics
of the Wigner logarithmic negativity can be studied. Interestingly,
time evolution of the Wigner logarithmic negativity (cf. Fig. \ref{fig:wigner-log})
for the PADFS evolving under photon loss channel shows similar behavior
as the variation of Wigner logarithmic negativity in noiseless case
with the displacement parameter (see Fig. \ref{fig:nonclassical-vol}).
This can be attributed to the fact that the photon loss channel is
a Gaussian channel and thus non-Gaussianity cannot increase due to
the application of this map. However, unlike Fig. \ref{fig:nonclassical-vol},
the nonclassicality and non-Gaussianity disappear beyond a threshold
value of $\kappa t$. Interestingly, this threshold value is closely
associated with a measure of nonclassicality proposed by Lee \citep{lee1991measure}. 

\section{Conclusion \label{sec:Result-and-discussion}}

Quantification of nonclassicality and quantum non-Gaussianity as well
as the study of the relation between them are problems of interest
in their own merit. These problems are addressed here by considering
PADFSs as a test case. Specifically, PADFS is chosen as it can be reduced
to various quantum states having important applications in different
limits. By construction, PADFSs are pure quantum states
and Figs. \ref{fig:nonclassical-vol} and \ref{fig:Single_photon}
clearly show that different choices of $n$ and $k$ lead to different
quantum states having different amount of non-Gaussianity. In this
work, apart from comparing non-Gaussianity, we have also used linear
entropy potential, Wigner logarithmic negativity and Wigner Yanase
skew information as measures of nonclassicality to compare the amount
of nonclasicality present in a set of different non-Gaussian pure
states. Here, it will be appropriate to note that the choice of 
nonclassicality measures is important as all the known measures of
nonclassicality cannot be used to compare the amount of nonclassicality
present in two non-Gaussian pure states. Specifically, nonclassical
depth, a well-known and rigorously studied measure of nonclassicality
introduced by Lee in \citep{lee1991measure} and subsequently used
by many authors (see \citep{miranowicz2015statistical,dodonov2003theory}
and references therein) cannot be used for this purpose as nonclassical
depth is 1 for all non-Gaussian pure states \citep{lutkenhaus1995nonclassical}.
Here, it will be apt to note that Lee used the fact that Gaussian
noise can deplete nonclassicality (as we observed in the dynamics
of Wigner function under photon loss channel) and considered the minimum
amount of noise needed to destroy nonclassicality as a measure of
nonclassicality which he referred to as nonclassical depth. Interestingly,
noise can change the nature of non-Gaussianity as discussed in \citep{genoni2013detecting}
as well as it can induce (classical) non-Gaussianity in the Gaussian
states due to phase fluctuations \citep{franzen2006experimental}.
The present study shows that quantum non-Gaussianity and nonclassicality
witnessed through Wigner negativity (and quantified using Wigner logarithmic
negativity) declines due to photon loss channels, which also infers
that the reported features can only be observed with highly efficient
detectors. However, we cannot discard from the present study that
the useful quantum non-Gaussianity (with Wigner negativity) of PADFS
is not eventually transformed into a non-Gaussian state having positive
Wigner function that cannot be used for quantum computing \citep{mari2012positive}.
Further, squeezing is expected to play an important role in combating
depleting Wigner negativity in view of some recent results \citep{filip2013gaussian}. 

In short, nonclassicality and non-Gaussianity present in the PADFS
is quantified, which shows that both photon addition and Fock parameter
enhance these quantum features where the former (latter) is a more
effective tool at small (large) displacement parameter. In contrast,
displacement parameter is observed to reduce the quantum features.
In view of Hudson's theorem and resource theory of non-Gaussianity
based on Wigner negativity \citep{albarelli2018resource}, the Gaussian
operations are free operations and thus, displacement operation and
photon loss channels are not expected to enhance non-Gaussianity and/or
Wigner negativity. Although, in case of pure PADFS, the Wigner negativity
captures all the quantum non-Gaussianity but cannot discard conclusively
the feasibility of quantum non-Gaussianty in PADFS measured through
inefficient detectors (or evolved through lossy channels) with the
positive Wigner function. Similarly, skew information based and linear
entropy potential measures show that Wigner function succeeds in detecting
all the nonclassicality of the pure PADFS. 

Results reported here seem to be experimentally realizable, as PACS,
which is a special case of PADFS has not only been prepared experimentally
by Bellini et al. \citep{zavatta2004quantum}, non-Gaussianity present
in this state is also measured experimentally \citep{barbieri2010non}.
Further, in \citep{malpani2019quantum}, a schematic diagram of the
setup that can generate PADFS is already reported. Keeping the above
and the recent progresses in quantum state engineering and quantum
information in mind, we conclude this article with an expectation
that PADFSs will soon be produced in laboratories and will be used
for performing quantum optical, meteorological and computing tasks.

\section*{Acknowledgment}

PM and AP acknowledge the support from DRDO, India project no. ANURAG/MMG/CARS/2018-19/071.
KT acknowledges the financial support from the Operational Programme
Research, Development and Education - European Regional Development
Fund project no. CZ.02.1.01/0.0/0.0/16\_019/0000754 of the Ministry
of Education, Youth and Sports of the Czech Republic.

\appendix

\section*{Appendix A: Details of Wigner function \label{sec:Appendix-A}}

It is straightforward to show that

\begin{equation}
\langle-\lambda|\rho|\lambda\rangle=\left|N\right|^{2}\frac{\left(-1\right)^{n+k}\lambda^{\star k}\lambda^{k}}{n!}\left(\lambda^{\star}+\alpha^{\star}\right)^{n}\left(\lambda-\alpha\right)^{n}\exp\left[\left(\alpha^{\star}\lambda-\alpha\lambda^{\star}\right)-\left(\left|\lambda\right|^{2}+\left|\alpha\right|^{2}\right)\right].\label{eq:A1}
\end{equation}
Substituting it in (\ref{eq:wigner-def}) and writing

\[
\begin{array}{lcl}
\left(\lambda^{\star}+\alpha^{\star}\right)^{n} & = & \sum_{t=0}^{n}{n \choose t}\left(\alpha^{\star}\right)^{n-t}\left(\lambda^{\star}\right)^{t},\\
\left(\lambda-\alpha\right)^{n} & = & \sum_{t^{\prime}=0}^{n}{n \choose t^{\prime}}\left(-\alpha\right)^{n-t^{\prime}}\left(\lambda\right)^{t^{\prime}},
\end{array}
\]
$W\left(\gamma,\gamma^{\star}\right)$ can be written in the form

\begin{equation}
\begin{array}{lcl}
W\left(\gamma,\gamma^{\star}\right) & = & \frac{2\left|N\right|^{2}}{\pi}\exp\left[2\left|\gamma\right|^{2}-\left|\alpha\right|^{2}\right]\frac{\left(-1\right)^{n+k}}{n!}\sum_{t=0}^{n}\sum_{t^{\prime}=0}^{n}{n \choose t}{n \choose t^{\prime}}\left(\alpha^{\star}\right)^{n-t}\left(-\alpha\right)^{n-t^{\prime}}\\
 & \times & \left(-\frac{\partial}{\partial\alpha}\right)^{k+t}\left(\frac{\partial}{\partial\alpha^{\star}}\right)^{k+t^{\prime}}\int\frac{d^{2}\lambda}{\pi}\exp\left[-\left|\lambda\right|^{2}-\eta^{\star}\lambda+\eta\lambda^{\star}\right],
\end{array}\label{eq:A2}
\end{equation}
where $\eta=2\gamma-\alpha,\,\frac{\partial}{\partial\alpha}\equiv-\frac{\partial}{\partial\eta},\,\frac{\partial}{\partial\alpha^{\star}}\equiv-\frac{\partial}{\partial\eta^{\star}}$.
Using the formula

\[
\int\frac{d^{2}z}{\pi}\exp\left[\zeta\left|z\right|^{2}+\xi z+\eta z^{\star}\right]=\frac{1}{\sqrt{\zeta^{2}}}\exp\left[-\frac{\zeta\xi\eta}{\zeta^{2}}\right],
\]
the integration over $\lambda$ yields $\exp\left[-\eta^{\star}\eta\right]$.
Further, setting $k+t=m^{\prime}$, $k+t^{\prime}=n^{\prime}$ , we
use the formula

\begin{equation}
\left(\frac{\partial}{\partial\eta}\right)^{m^{\prime}}\left(\frac{\partial}{\partial\eta^{\star}}\right)^{n^{\prime}}\exp\left[-\eta^{\star}\eta\right]=\begin{cases}
\left(-1\right)^{n^{\prime}}m^{\prime}!\exp\left[-\eta\eta^{\star}\right]\eta^{n^{\prime}-m^{\prime}}L_{m^{\prime}}^{n^{\prime}-m^{\prime}}\left(\eta\eta^{\star}\right) & m^{\prime}\leq n^{\prime},\\
\left(-1\right)^{m^{\prime}}n^{\prime}!\exp\left[-\eta\eta^{\star}\right]\eta^{\star\left(m^{\prime}-n^{\prime}\right)}L_{n^{\prime}}^{m^{\prime}-n^{\prime}}\left(\eta\eta^{\star}\right) & n^{\prime}\leq m^{\prime}.
\end{cases}\label{eq:A4}
\end{equation}

Substituting in (\ref{eq:A2}), the double sum can be re-ordered to
give the final expression (\ref{eq:Wigner}) for the Wigner function
$W\left(\gamma,\gamma^{\star}\right).$ 
\end{document}